%% file: main.tex
\pdfoutput=1

\documentclass[3p,onecolumn,10pt,times]{elsarticle}

\usepackage{amssymb}
\include{0_preamble}

\journal{Computer-Aided Design, SI: AI + Design}

\begin{document}

\begin{frontmatter}

% \title{PILL-ID: Inverse Design of Multi-Material Polypills for Prescribed Drug-Release Kinetics}
\title{PILL-CoDe: Inverse Design of Polypills via Automatic Differentiation for Prescribed Drug-Release Kinetics}
\author[UW]{Rahul Kumar Padhy}
\author[NU]{Aaditya Chandrasekhar}
\author[KU]{Amir M. Mirzendehdel\corref{cor1}}

\affiliation[UW]{organization={Department of Mechanical Engineering, University of Wisconsin-Madison},%Department and Organization
            city={Madison},
            state={WI},
            country={USA}}

\affiliation[NU]{organization={Department of Mechanical Engineering, Northwestern University},%Department and Organization
city={Evanston},
state={IL},
country={USA}}
\affiliation[KU]{organization={Department of Aerospace Engineering, University of Kansas},%Department and Organization
            city={Lawrence},
            state={KS},
            country={USA}}

\cortext[cor1]{Corresponding author}

\input{1_abstract}


% \begin{keyword}
% %% keywords here, in the form: keyword \sep keyword
% Polypills \sep Personalized medication \sep Neural network \sep Inverse design \sep Additive manufacturing
% \end{keyword}

\end{frontmatter}

 % \linenumbers

\input{2_introduction}
\input{3_method}
\input{4_results}

\input{5_conclusions}

% \section*{Author Contributions}
% R.K.P: Conceptualization, Methodology, Software, Visualization, Writing-original draft, Writing–review and editing; A.C: Conceptualization, Software, Writing–review and editing; A.M: Conceptualization, Methodology, Writing-original draft, Writing–review and editing

\section*{Data Availability and Replication of Results}
The Python code is available at{\href{https://github.com/DEL-KU/PILL-CoDe}{https://github.com/DEL-KU/PILL-CoDe}}

\section*{Declaration of Competing Interest}
The authors declare that they have no known competing financial interests or personal relationships that could have appeared to influence the work reported in this paper.

\section*{Declaration of generative AI and AI-assisted technologies in the manuscript preparation process}
During the preparation of this manuscript, the authors used Google’s Gemini, OpenAI’s ChatGPT{, and Anthropic's Claude} to improve the
language and readability. After using these tools, the authors reviewed and edited the content as needed and take full
responsibility for the content of the publication.

% The acknowledgements will be added in future revisions, to respect the anonymous review process.

% \appendix

 \bibliographystyle{elsarticle-num} 
 \bibliography{references}





% \end{thebibliography}
\end{document}

%% file: 0_preamble.tex
\usepackage{amsmath}
\usepackage{empheq}

\usepackage{hyperref}

\usepackage{lineno}
\usepackage{hyperref}
\usepackage{enumitem}
\modulolinenumbers[5]

\usepackage{subcaption}
\usepackage{mathtools}
\usepackage{amsmath}
\usepackage{amssymb}
\usepackage{stmaryrd}
\usepackage{amsthm}
\usepackage{amsfonts}
\usepackage{dsfont}
\usepackage{graphicx}
\usepackage{upgreek}
\usepackage{bm}
\usepackage{color}
\usepackage{float}
\usepackage{tikz-cd}
\usepackage{scrextend}
\usepackage[cal=boondoxo]{mathalfa}
\usepackage{nicefrac}
\usepackage{algorithm,algpseudocode} 
\usepackage{cleveref}
\usepackage{soul}

\usepackage[colorinlistoftodos]{todonotes}
\usepackage{tcolorbox}
\usepackage{booktabs}
\usepackage{soul}

\biboptions{numbers,sort&compress}

\DeclareFontFamily{OT1}{pzc}{}
\DeclareFontShape{OT1}{pzc}{m}{it}{<-> s * [1.200] pzcmi7t}{}
\DeclareMathAlphabet{\mathpzc}{OT1}{pzc}{m}{it}

\theoremstyle{definition}

\newcommand{\com}[1]{} %Use \com{...} to commenting out multiple paragraphs

%% file: 1_abstract.tex
\begin{abstract}
{Polypills are single oral dosage forms that combine multiple active pharmaceutical ingredients and excipients, enabling fixed-dose combination therapies, coordinated multi-phase release, and precise customization of patient-specific treatment protocols. Recent advances in additive manufacturing facilitate the physical realization of multi-material excipients, offering superior customization of target release profiles. However, polypill formulations remain tuned by ad hoc parameter sweeps. The current design workflows are ill-suited for the systematic exploration of the high-dimensional space of shapes, compositions, and release behaviors.}

{We present PILL-CoDe, a polypill co-design framework that simultaneously optimizes tablet geometry and excipient distribution to match prescribed drug-release kinetics. The framework couples a supershape parametrization of the pill geometry with a coordinate-based neural network representation of the excipient distribution, and governs dissolution through a coupled system of modified Allen-Cahn and Fickian diffusion equations. Implemented in JAX, the entire pipeline is end-to-end differentiable, with automatic differentiation providing exact sensitivities for gradient-based co-optimization of shape and composition under manufacturability constraints. We demonstrate the method through single-phase and multi-excipient case studies, showing accurate matching of both monotonic and non-monotonic target release profiles.}
\end{abstract}
% \begin{abstract}

% Polypills are single oral dosage forms that combine multiple active pharmaceutical ingredients and excipients, enabling fixed-dose combination therapies, coordinated multi-phase release, and precise customization of patient-specific treatment protocols. Recent advances in additive manufacturing facilitate the physical realization of multi-material excipients, offering superior customization of target release profiles. However, polypill formulations remain tuned by ad hoc parameter sweeps; this reliance renders current design workflows ill-suited for the systematic exploration of the high-dimensional space of shapes, compositions, and release behaviors.

% We present an automated design framework for polypills that leverages topology optimization to match dissolution behaviors with prescribed drug release kinetics. In particular, we employ a supershape parametrization to define geometry/phase distribution, a neural network representation to specify excipient distribution, and a coupled system of modified Allen–Cahn and Fick's diffusion equations to govern dissolution kinetics. The framework is implemented in JAX, utilizing automatic differentiation to compute sensitivities for the co-optimization of pill shape and constituent distribution. We validate the method through single-phase and multi-excipient case studies.

% \end{abstract}

%% file: 2_introduction.tex
\section{Introduction}
\label{sec:intro}

\begin{figure}[!h]
  \centering
  \includegraphics[scale=0.45]{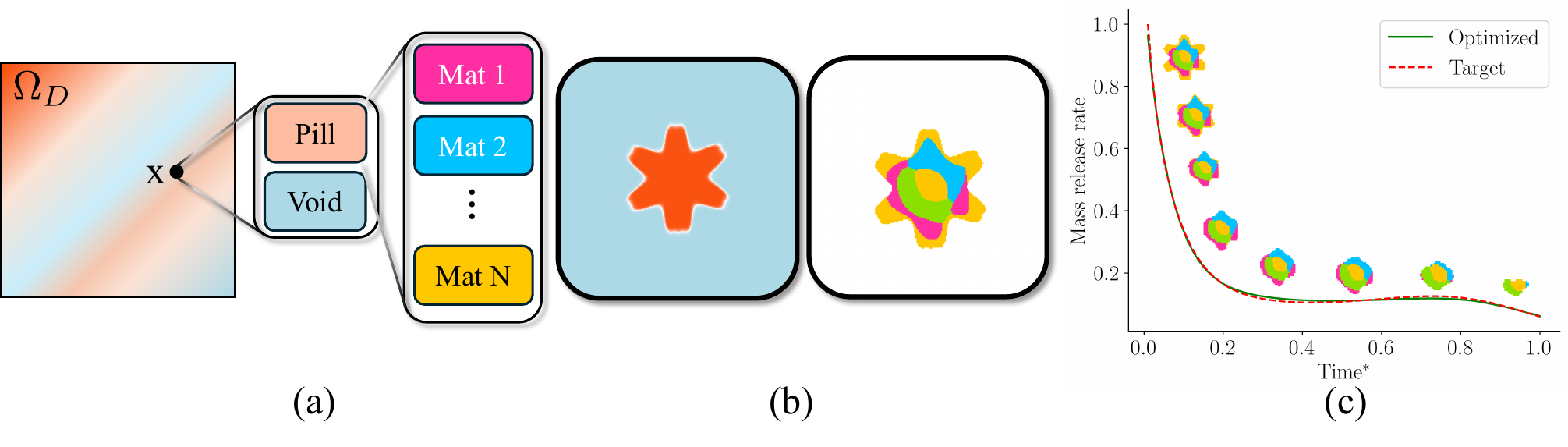}
  \caption{Graphical abstract: Given (a) a design domain and a set of candidate materials, the framework employs multimaterial {inverse design} governed by phase-field dissolution kinetics to tailor the polypill release profile. By coupling a supershape parametrized geometry with a neural material distribution, the framework co-optimizes (b) the geometry and internal material distribution, yielding (c) a design that matches the target release curve.}
  \label{fig:overview}
\end{figure}

Personalized multidrug tablets, often referred to as \textit{polypills}, enable patient-specific release profile by combining multiple active pharmaceutical ingredients and excipients \cite{yasin2024fabrication}. Simultaneously, Additive manufacturing  of pills has emerged as a promising approach for creating tailored dosages over time through direct specification of geometry, internal architecture, and material distribution from digital designs. Furthermore, additive manufacturing enables point-of-care production to improve supply-chain resilience and quality of care by reducing lead times \cite{FDA2022DM,FDA2023Stakeholder}. {Five additive technologies dominate pharmaceutical research: binder jet printing~\cite{sen2021pharmaceutical}, fused deposition modeling~\cite{iqbal2024status}, semi-solid extrusion~\cite{seoane2021semi}, selective laser sintering ~\cite{charoo2021opportunities}, and stereolithography~\cite{deshmane2021stereolithography,Xu2020_IJPharm}. Aprecia's ZipDose\texttrademark{} platform (2008) first showed binder jet printing could produce orally disintegrating, high-dose formulations~\cite{FierceBiotech2015}, and parallel extrusion- and inkjet-based studies established that geometry and internal architecture can be tuned to modulate drug-release kinetics~\cite{scoutaris2011inkjet, trenfield20183d}, a principle reinforced by the 2015 FDA approval of \textit{Spritam} (levetiracetam) as the first 3D-printed tablet~\cite{Aprecia2015}. Building on this foundation, reconfigurable ``chemistry-on-demand'' systems~\cite{adamo2016demand, zhang2018advanced}, defense and regulatory initiatives toward point-of-care production~\cite{DARPA_Battlefield, lewin2016pharmacy, DARPA_BioMOD_BAA2012, FDA2022DM, FDA2023Stakeholder}, and space-based in-situ manufacturing efforts~\cite{ISS_InSPA, NASA_InSPA} together signal a broader shift toward distributed, on-demand pharmaceutical production.}

{The ability to prescribe release kinetics depends on both the external geometry and the internal material arrangement of the dosage form. Geometry controls the exposed surface area and the evolving diffusion length during dissolution, while the placement of excipients and active ingredients determines where fast- and slow-dissolving regions are encountered as the tablet erodes. Experimental studies in pharmaceutical additive manufacturing have shown that changing tablet shape, infill architecture, shell thickness, and material composition can substantially alter dissolution and release behavior~\cite{goyanes2015effect, goyanes2016fused, khaled20153d, sadia2018channelled}. These observations suggest that geometry-only design is intrinsically limited: it can modulate release through surface exposure, but it cannot independently prescribe when different ingredients or excipient-controlled dissolution rates become active. Co-design therefore expands the reachable set of release profiles by coupling shape-controlled interface evolution with material-controlled local dissolution kinetics.}

{Despite recent technical advances and emerging regulatory interest, systematic design tools for efficiently exploring the geometry and material layout of multi-ingredient tablets remain limited~\cite{konta2017personalised}. Three limitations are particularly relevant.}

{First, existing inverse-design methods do not jointly optimize tablet geometry and multi-ingredient composition. Panetta et al.~\cite{Panetta2022TOG} introduced a ``shape-from-release'' framework that couples a differentiable dissolution model with PDE-constrained inverse design and fabrication-aware regularization, demonstrating close agreement between simulated and measured release curves for single-material, 3D-printed structures. Altunay et al.~\cite{altunay2025computational} extended this idea to multi-material tablets by optimizing the spatial distribution of a single active ingredient, carried by two materials, within a fixed capsule-shaped geometry to match prescribed release profiles under uncertainty. However, neither approach co-optimizes tablet geometry together with the spatially varying composition of multiple active ingredients, which is central to polypill design. This distinction is important because geometry primarily changes the time evolution of exposed surface area, whereas material placement determines which local dissolution rates and ingredients are exposed at different stages of erosion. Our formulation treats both geometry and per-voxel composition of multiple active ingredients and excipients as coupled design variables.}

{Second, prior formulations rely on simplified geometric dissolution models rather than coupled dissolution-diffusion transport. Existing inverse-design approaches idealize release as a front-propagation problem governed by an Eikonal equation, where local slowness or surface velocity is prescribed algebraically from the design density or material fractions~\cite{Panetta2022TOG,altunay2025computational}. This approximation is computationally attractive and can be effective when dissolution is well stirred or reaction limited. However, it does not explicitly resolve dissolved-drug concentration fields, local concentration buildup near the dissolving interface, saturation effects, or diffusion-limited transport through the surrounding medium. These effects are central to classical dissolution theory, including Noyes-Whitney and Nernst-Brunner type models, where the dissolution rate depends on the difference between saturation and local concentration and on diffusive transport away from the interface~\cite{dokoumetzidis2006century, siepmann2013mathematical}. To capture these mechanisms, we model the evolving tablet interface with a modified Allen-Cahn phase-field equation and couple it to Fickian diffusion of the dissolved species. While direct experimental validation of Allen-Cahn phase-field dissolution models for pharmaceutical tablets remains limited, phase-field formulations have been used and validated for moving-interface dissolution and precipitation problems in related transport settings~\cite{xu2008phase,xu2012phase,yang2021numerical,bringedal2020phase}. In the present work, this formulation provides a differentiable diffuse-interface model of pharmaceutical dissolution that is more physically expressive than prescribed front-speed models, while remaining suitable for gradient-based inverse design.}

{Third, there remains a gap between physics-based inverse design (e.g., topology optimization~\cite{bendsoe2013topology}) and controlled-release tablet design. Physics-informed inverse design has been widely used to design structures, fluids, heat-transfer systems, and multi-material layouts by embedding governing equations directly inside the optimization loop~\cite{sigmund2013topology, alexandersen2020review}. This paradigm is particularly well suited to controlled-release inverse design because the desired performance is not a static geometric feature, but a time-dependent functional response generated by coupled transport, moving interfaces, and material heterogeneity. However, applying this idea to richer dissolution physics is challenging because hand-derived adjoint sensitivities become increasingly cumbersome as the forward model gains physical fidelity. We address this barrier by using automatic differentiation through the discretized forward solver, together with implicit differentiation of the nonlinear solve, to obtain exact discrete sensitivities for the coupled dissolution-diffusion model. This makes the co-design of tablet geometry and multi-ingredient composition tractable under a physics-informed inverse-design framework.}

{As mentioned above, existing methods model dissolution as a geometric  front-propagation process, which is effective under well-stirred, reaction-limited conditions \cite{Panetta2022TOG} but does not explicitly resolve the coupled dissolution-diffusion kinetics that can govern drug release in multi-material systems where local concentration buildup modulates the dissolution 
rate \cite{sleziona2022modeling, sleziona2021determination}. \cite{altunay2025computational} note that incorporating such coupling ``\emph{would require significantly more numerical side work to define the adjoint equation for the gradients of the objective functions.}''} In this paper, we address this limitation by formulating a differentiable, coupled dissolution–diffusion model and integrating it into a gradient-based optimization framework with automated sensitivity analysis, thereby enabling the systematic co-design of geometry and spatially varying compositions under {under a coupled dissolution-diffusion formulation}.

{More specifically, the contributions of this paper are:}
\begin{enumerate}
    \item {We introduce a hybrid design parameterization in which supershapes define the global tablet geometry to enforce a connected, monolithic dosage form, while coordinate-based neural networks encode the internal multi-material distribution with high spatial flexibility.}
    {\item We develop a differentiable dissolution-diffusion simulator that couples a modified Allen-Cahn phase-field model for interface evolution with Fickian diffusion of dissolved species, allowing local concentration buildup and saturation effects to influence release kinetics.}
    {\item We formulate a physics-informed inverse design framework for controlled-release of active ingredients, where the governing dissolution-diffusion equations are embedded directly in the optimization loop and sensitivities are computed automatically through the discretized solver.}
    {\item We demonstrate that co-designing geometry and excipient distribution expands the reachable set of release profiles compared with geometry-only optimization, enabling accurate matching of both monotonic and non-monotonic instantaneous release-rate profiles.}
\end{enumerate}

%% file: 3_method.tex
\section{Proposed Method} 
\label{sec:method}

This study addresses the concurrent optimization of polypill geometry and constituent material distribution to achieve prescribed drug release profiles. We begin by assuming a  computational domain $\Omega$ subject to boundary conditions governing pill dissolution. The polypill is modeled as a heterogeneous carrier matrix composed of multiple spatially distributed excipients embedding an active ingredient \cite{van2020role, yasin2024fabrication, khaled20153d}. The excipients function as medically inert materials that govern the dissolution kinetics, thereby controlling the drug release profile while the aggregate dosage remains fixed by the prescribed active ingredient concentration \cite{van2020role, saylor2007diffuse}. We assume the initial active ingredient concentration is uniform throughout the solid domain, while the surrounding solvent is initially devoid of any active ingredient. Furthermore, we assume the availability of $S$ distinct excipients, each characterized by a unique dissolution constant. The excipient material distribution is initialized uniformly with equal volume fractions for each excipient.

\begin{figure}[!h]
  \centering
  \includegraphics[scale=0.35]{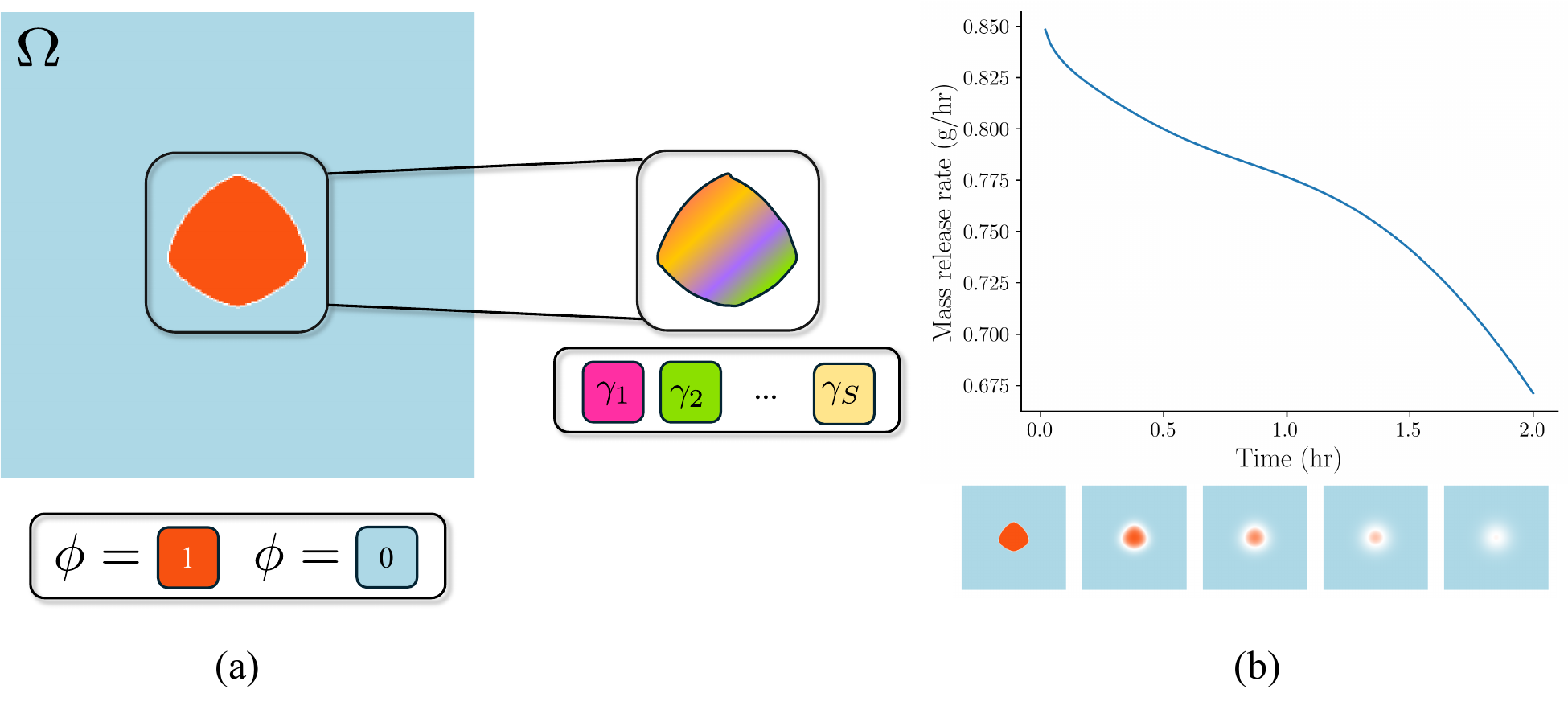}
  \caption{(a) Given a computational domain $\Omega$, we optimize the pill shape (region where $\phi = 1$) and the distribution of excipients $(\gamma_1,\ldots,\gamma_S)$ so that the resulting dissolution kinetics match the prescribed target shown in (b).}
  \label{fig:drug_target}
\end{figure}

The central premise of this work is that the drug release profile is governed by two design factors: first, the pill geometry, which dictates the evolving surface area available for solvent interaction \cite{mohseni2023tablet, goyanes2015effect}; and second, the spatial distribution of excipients \cite{van2020role}, which modulates the local dissolution rate at the solid-solvent interface. Consequently, the design objective is to determine the optimal {pill geometry} and spatial material distribution (\Cref{fig:drug_target}(a)) to achieve prescribed target dissolution kinetics (\Cref{fig:drug_target}(b)).

To this end, we propose a multi-material {inverse design} framework that simultaneously optimizes the shape of the pill and the excipient material distribution. Our goal is to maximize design space to match complex kinetics, while enforcing topological connectivity required for manufacturability. In particular, we use supershapes \cite{gielis2003generic, padhy2024tomas} to parameterize the exterior shape (\Cref{sec:method_designRep_phaseField}), ensuring a connected design, while a coordinate-based NN represents \cite{chandrasekhar2021multi} the excipient distribution to maximize design freedom (\Cref{sec:method_designRep_NN}). Furthermore, the dissolution process is governed by a modified Allen-Cahn phase-field equation \cite{allen1979microscopic, xu2012phase}, coupled with a Fickian model for diffusion \cite{fick1855ueber, goodacre1981mathematical}. 

We begin by detailing the governing physics and corresponding finite element formulations in \Cref{sec:method_govEq} and \Cref{sec:method_fea}, respectively. We then describe the hybrid design representation in \Cref{sec:method_designRep}. This is followed by the derivation of effective material properties from the phase and excipient distributions in \Cref{sec:method_materialModel}. Finally, \Cref{sec:method_optimization} presents the optimization formulation, including the objective function, constraints, loss function, sensitivity analysis, and the numerical optimization procedure.

%----------------------------------%
\subsection{Governing Equations}
\label{sec:method_govEq}

The coupled dissolution and diffusion kinetics are modeled within a computational domain $\Omega$, which encompasses both the dissolving polypill and the surrounding solvent medium. The state of the system is defined by two field variables: the phase-field variable $\phi(\bm{x}, t) \in [0, 1]$, and the concentration field $C(\bm{x}, t)$. The phase-field variable serves as an indicator function, where $\phi = 1$ denotes the solid pill domain (i.e., excipient and active ingredient), $\phi = 0$ denotes the solvent, and intermediate values represent the diffuse solid-fluid interface. The concentration field $C(\bm{x}, t)$ represents the mass of the dissolved solute per unit volume.

We assume the domain is initialized with a uniform concentration $C(\bm{x}, 0) = 0$ (\Cref{eq:residual_form_c_ic}), corresponding to an undissolved pill and a solvent initially devoid of solute. The initial phase field $\phi(\bm{x}, 0)$ defines the initial geometry of the polypill $\phi_0(\bm{x})$ (\Cref{eq:residual_form_phi_ic}), which is optimized. To model the dissolution process, we assume an infinite sink condition for the solvent boundaries to prevent saturation \cite{siepmann2013mathematical}. Consequently, we enforce a zero-Dirichlet boundary condition for the concentration (\Cref{eq:residual_form_c_bc}) and a zero-flux Neumann condition (\Cref{eq:residual_form_phi_bc}) for the phase field along outward normal $\bm{n}$ on the domain boundary $\Gamma$.

The evolution of the pill interface is governed by a modified Allen-Cahn equation \cite{xu2008phase, yang2021numerical, singer2008phase}, augmented with a dissolution forcing term derived from Nernst-Brunner kinetics \cite{siepmann2013mathematical, goodacre1981mathematical}. The governing dissolution equation can then be expressed as:

\begin{subequations}
\label{eq:residual_form_phi}
\begin{empheq}[left={R_{\phi}(\phi, C) \coloneqq \empheqlbrace}]{align}
  \frac{\partial \phi}{\partial t}
  + M_{\phi} \left( \psi'(\phi) - W \epsilon_t^2 \nabla^2 \phi \right)
  - f_{\text{diss}} &= 0,
  && \text{for } \bm{x} \in \Omega,\ \forall t > 0,
  \label{eq:residual_form_phi_pde} \\[0.3em]
  \nabla \phi \cdot \bm{n} &= 0,
  && \text{for } \bm{x} \in \Gamma,\ \forall t,
  \label{eq:residual_form_phi_bc} \\[0.3em]
  \phi(\bm{x}, 0) - \phi_0(\bm{x}) &= 0,
  && \text{for } \bm{x} \in \Omega,
  \label{eq:residual_form_phi_ic}
\end{empheq}
\end{subequations}

where $M_{\phi}$ is the interface mobility, $W$ is the double-well potential barrier height, and $\epsilon_t$ is the interface thickness parameter. The potential density function is defined as $\psi(\phi) = W\phi^2(1 - \phi)^2$ \cite{yang2024melting}, ensuring stable minima at the pure phases. The dissolution forcing term, $f_{diss}$, couples the phase evolution to the concentration field $C$ \cite{sleziona2022modeling, siepmann2013mathematical, bringedal2020phase}:
\begin{equation}
    f_{diss} = -\frac{k}{\rho_{s}} (C_{sat} - C) |\nabla \phi|
    \label{eq:forcing}
\end{equation}
Here, $k$ is the dissolution rate constant, $\rho_{s}$ is the solid density, and $C_{sat}$ is the saturation solubility. The term $|\nabla \phi|$ localizes the dissolution driving force to the solid-solvent interface. {The concentration-dependent factor $(C_{\mathrm{sat}} - C)$ reduces the local dissolution rate as the solute concentration approaches saturation near the interface. This is consistent with classical dissolution theory, in which the overall dissolution rate depends on the relative rates of interfacial and transport processes, with diffusion through the unstirred region 
often playing a rate-controlling role \cite{siepmann2013mathematical, sleziona2021determination, 
mattusch2025intrinsic}.}

Simultaneously, the transport of the dissolved solute is governed by the diffusion equation \cite{fick1855ueber, xu2008phase, bringedal2020phase} (neglecting any advection effects), can be expressed as:

\begin{subequations}
\label{eq:residual_form_c}
\begin{empheq}[left={R_{c}(\phi, C) \coloneqq \empheqlbrace}]{align}
  \frac{\partial C}{\partial t}
  - \nabla \cdot \big(D(\phi) \nabla C\big)
  - S_{\text{source}} &= 0,
  && \text{for } \bm{x} \in \Omega,\ \forall t > 0,
  \label{eq:residual_form_c_pde} \\[0.3em]
  C &= 0,
  && \text{for } \bm{x} \in \Gamma,\ \forall t,
  \label{eq:residual_form_c_bc} \\[0.3em]
  C(\bm{x}, 0) &= 0,
  && \text{for } \bm{x} \in \Omega,
  \label{eq:residual_form_c_ic}
\end{empheq}
\end{subequations}

where $D$ is the diffusion coefficient, and the source term $S_{source}$ accounts for the mass transfer from the solid phase to the solution, ensuring mass conservation \cite{dokoumetzidis2006century, siepmann2013mathematical, sleziona2022modeling}:
\begin{equation}
    S_{source} = k (C_{sat} - C) |\nabla \phi|
\end{equation}

{ Furthermore, the concentration boundary condition models an idealized \textit{in vitro} sink-condition setting. Thus, the concentration field may evolve within the domain and develop local gradients near the dissolving interface, while the surrounding medium acts as an ideal sink.}

The coupled system of residuals $R_{\phi}$ and $R_{c}$ is spatially discretized and solved using standard finite element procedures as detailed in Section \ref{sec:method_fea}. 

%----------------------------------%
\subsection{Design Representation} 
\label{sec:method_designRep}

Having established the governing equations, we now define the design parameterization strategy. Recall that the optimization objective is two-fold: (a) to determine the overall shape of the polypill, and (b) to optimize the spatial distribution of excipients within the solid domain.

To ensure the optimized design constitutes a single, connected pill rather than fragmented collection of particulates, we require that the pill designed be a topologically connected domain. Simultaneously, to maximize the tailorability of the release profile, we seek to promote material heterogeneity within the pill's interior.

Consequently, we adopt a hybrid parameterization strategy. In particular, we utilize supershapes to define the global {geometry}, ensuring a monolithic structure (\Cref{sec:method_designRep_phaseField}), while employing coordinate-based neural networks to represent the internal material distribution, thereby enabling {high resolution design representation independent of mesh resolution}. (\Cref{sec:method_designRep_NN}).

%----------------------------------%
\subsubsection{Phase Field using Supershapes} \label{sec:method_designRep_phaseField}

To enforce topological connectivity and avoid domain fragmentation, we employ a geometric parameterization based on supershapes (Gielis curves) \cite{gielis2003generic, padhy2024tomas}. Defined as a generalization of super-quadrics \cite{gielis2003generic}, this formulation provides a versatile parametrization for generating a broad spectrum of continuous geometries, spanning from standard convex forms (e.g., capsules) to complex non-convex (e.g., multi-lobed) structures, using a compact set of parameters. 

We define the pill geometry in a local polar coordinate system $(r, \varphi)$. The radial boundary distance, $R(\varphi)$, is governed by the reduced-order Gielis equation:

\begin{equation}
    R(\varphi) = \left[ \left| \frac{1}{a} \cos \left( \frac{m \varphi}{4} \right) \right|^{n} + \left| \frac{1}{b} \sin \left( \frac{m \varphi}{4} \right) \right|^{n} \right]^{-\frac{1}{n}}
    \label{eq:supershape_radius}
\end{equation}

\begin{figure}[!h]
  \centering
  \includegraphics[scale=0.4]{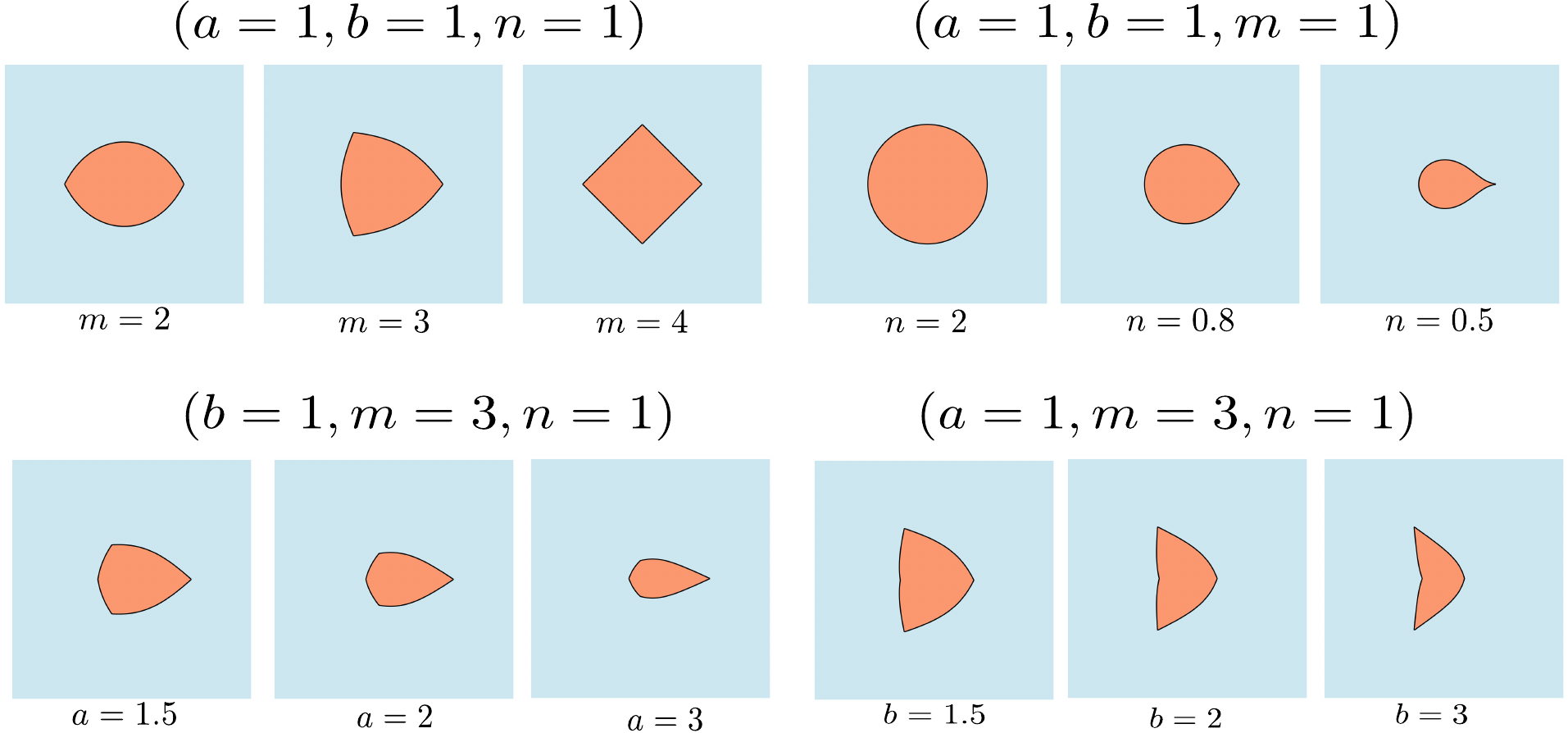}
  \caption{Variations in the supershape's parameters.}
\label{fig:superellipse_param_variations}
\end{figure}

Here, $a$ and $b$ denote the semi-axis scaling factors, $m$ controls the rotational symmetry (number of lobes), and $n$ is the curvature exponent related to the squareness parameter (\Cref{fig:superellipse_param_variations}).

Furthermore, to enable arbitrary positioning and orientation within the computational domain, we apply an affine transformation to the global coordinates $\bm{x}$. In particular, we let $\bm{x}_c = [c_x, c_y]^\top$ denote the translation vector and $\theta$ denote the rotation angle. The local coordinates $\bm{x}_{loc}$ used to evaluate the polar parameters $(r, \varphi)$ can then be obtained via (\Cref{eq:affine_transform}):

\begin{equation}
    \bm{x}_{loc} = \bm{R}(\theta) (\bm{x} - \bm{x}_c)
    \label{eq:affine_transform}
\end{equation}

where $\bm{R}(\theta)$ is the standard 2D rotation matrix.

To map this implicit shape onto the finite element mesh for the dissolution simulation (\Cref{sec:method_fea}), we begin by defining an radial distance field $\Phi(\bm{x})$. Recall that strict Euclidean signed distance fields (SDF) are often used in feature mapping methods in TO \cite{kumar2025treetop, padhy2025photos}. We instead utilize an approximation that preserves the zero-level set boundary and differentiability, rendering it sufficient for our inverse-design formulation. This field is defined as:

\begin{equation} 
\Phi(\bm{x}) = r - R(\varphi) \label{eq:distance_field_supershape} 
\end{equation}

where $r = \|\bm{x}_{loc}\|$ and $\varphi = \arctan(y_{loc} / x_{loc})$. This formulation ensures that $\Phi < 0$ inside the pill, $\Phi > 0$ outside, and $\Phi = 0$ on the boundary. {This approximation is computationally simpler than recomputing a true signed distance field after each geometry update, while preserving the  zero-level set of the supershape.}

\begin{figure}[!h]
  \centering
  \includegraphics[scale=0.5]{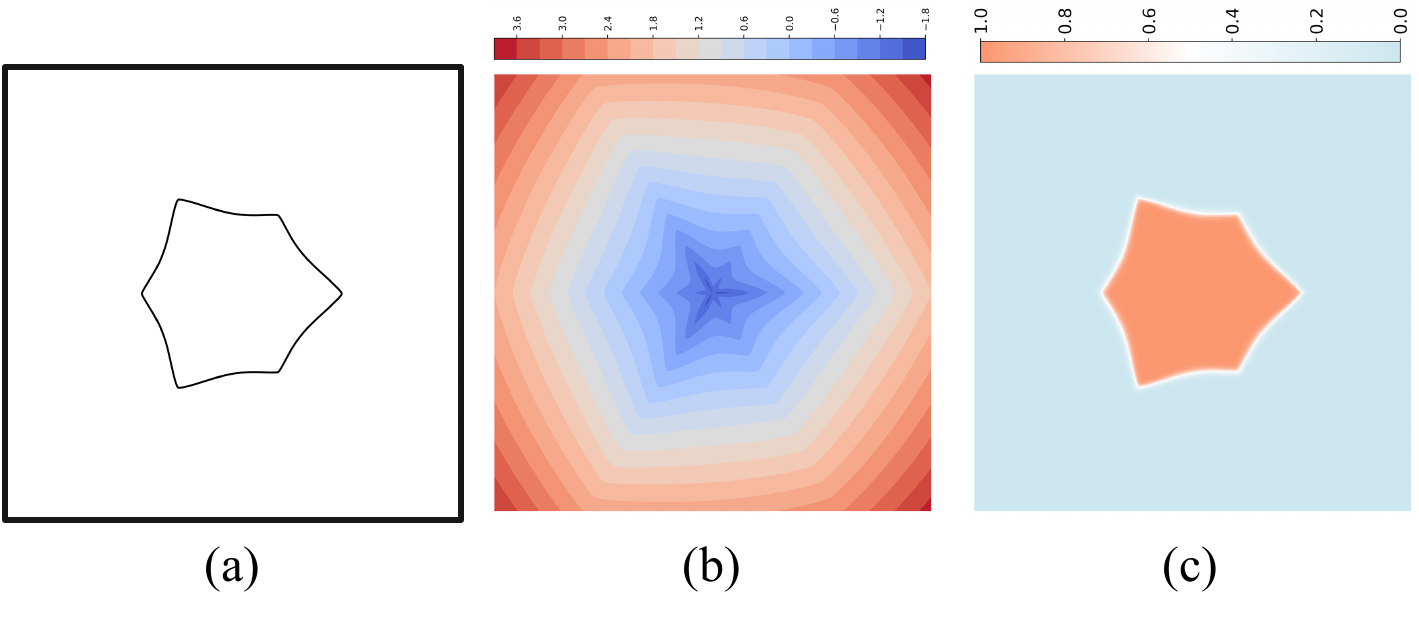}
  \caption{(a) The supershape as defined by its parameters ($a=1.8$, $b=1.5$, $n=1.25$, $m=6$). (b) The radial distance field. (c) The phase-field variable obtained upon projecting the distance field.}
  \label{fig:superellipse_shape}
\end{figure}

Finally, to obtain a differentiable approximation of the phase-field variable $\hat{\phi}(\bm{x})$, we project the radial distance field into a smooth density field using a hyperbolic tangent projection (\Cref{fig:superellipse_shape}) \cite{wang2011projection}:

\begin{equation}
\phi(\bm{x}) = \frac{1}{2} \left[ 1 - \tanh \left( \frac{\Phi(\bm{x})}{\mu} \right) \right] \label{eq:density_projection} \end{equation}

here, $\mu$ governs the steepness of the solid-solvent interface; in our experiments, we set $\mu = 10^{-4}$. This projection maps $\phi \approx 1$ to the material interior (where $\Phi < 0$) and $\phi \approx 0$ to the exterior solvent (where $\Phi > 0$).

 Finally, we note that the configuration of the pill is described by a set of seven parameters, $\bm{\zeta} \in \mathbb{R}^7$, comprising the translation coordinates ($c_x, c_y$), the orientation angle ($\theta$), the semi-axis scaling factors ($a, b$), the squareness parameter ($n$), and the rotational symmetry order ($m$). Collectively, the supershape design vector can be defined as in \Cref{eq:geo_params} which is further subject to optimization,

\begin{equation}
\bm{\zeta} = [c_x, c_y, \theta, a, b, n, m]^\top
\label{eq:geo_params}
\end{equation}

%----------------------------------%
\subsubsection{Material Distribution using Neural Networks}
\label{sec:method_designRep_NN}

Having obtained the pill's {geometry}, we now define the material distribution of the excipients. In particular, we let $\gamma_s(\bm{x})$ denote the presence of the $s$-th excipient at spatial coordinate $\bm{x}$, where $s \in \{1, \dots, S\}$. While a physically realizable design requires a discrete material assignment $\gamma_s(\bm{x}) \in \{0, 1\}$, we relax this condition for continuous gradient-based optimization, allowing $0 \leq \gamma_s(\bm{x}) \leq 1$. The material composition at any point is thus defined by the vector $\bm{\gamma}(\bm{x}) = [\gamma_1(\bm{x}), \dots, \gamma_S(\bm{x})]^\top$, subject to the partition of unity constraint $\sum\limits_{s=1}^{S} \gamma_s(\bm{x}) = 1$.

\begin{figure}[!h]
  \centering
  \includegraphics[scale=0.65]{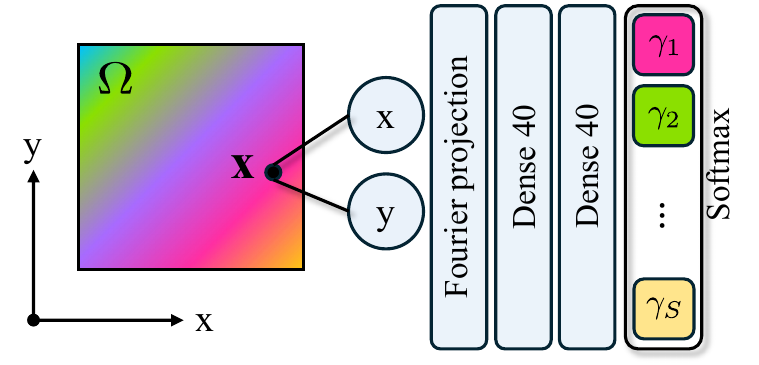}
  \caption{The neural network maps spatial coordinates to the material distribution.}
  \label{fig:nn_architecture}
\end{figure}

To parameterize this spatially varying material field, we employ a coordinate-based neural network (\Cref{fig:nn_architecture}) {\cite{chandrasekhar2021multi, chandrasekhar2022approximate, chandrasekhar2021tounn}}. Unlike conventional TO approaches where design variables are explicitly defined on the finite element discretization, this neural representation offers several distinct advantages for multi-material design: (a) the network architecture naturally enforces the partition of unity constraint via its output activation; (b) the design is decoupled from the simulation mesh, allowing for a compact set of design variables (network weights) independent of mesh resolution; (c) the material field is analytically defined everywhere, facilitating the recovery of crisp, high-resolution designs during post-processing; and (d) the analytic definition enables the precise computation of spatial gradients $\nabla_{\bm{x}} \bm{\gamma}$ via back-propagation, which is critical for obtaining sharp material interfaces.

The proposed network architecture consists of the following components:

\begin{enumerate}
    \item \textbf{Input Layer:} The network accepts the spatial coordinates $\bm{x} \in \mathbb{R}^d$ ($d=2$ in 2-Dimensions) as input.
    
    \item \textbf{Fourier Projection:} The Euclidean coordinates are mapped to a higher-dimensional feature space using a Fourier projection layer \cite{tancik2020fourier}. This (a) mitigates the spectral bias of standard multi-layer perceptrons, (b) enables the representation of high-frequency spatial variations, and (c) promotes faster convergence \cite{chandrasekhar2022approximate}.

    \item \textbf{Hidden Layers:} The feature vector is processed through a series of two dense hidden layers (with 40 neurons each) with ReLU activation functions \cite{agarap2018deep}.
    
    \item \textbf{Output Layer:} The final layer consists of $S$ neurons corresponding to the excipient volume fractions. A Softmax activation function is applied to the output, intrinsically satisfying the requirements $\sum\limits_{s=1}^S \gamma_s = 1$ and $0 \le \gamma_s \le 1$.
\end{enumerate}

The trainable weights and biases $\bm{w}$ of this network constitute the primary design variables for the material distribution. The weights  are initialized using the Xavier normal scheme (\cite{glorot2010understanding}) and subsequently updated to yield a spatially uniform material distribution, assigning equal volume fractions ($1/S$) to each constituent excipient phase. 

%----------------------------------%

\subsection{Material Model}
\label{sec:method_materialModel}

With the pill {geometry} ($\bm{\phi} = \left[ \phi(\bm{x}) \right]$) and the internal excipient distribution ($\bm{\gamma}= \left[ \bm{\gamma}(\bm{x}) \right]$) established, we now define the constitutive relations governing the local material properties. To model the effective diffusivity across the domain, we employ a smooth interpolation between the solvent and solid phases. The diffusion coefficient $D(\phi)$ is defined as:

\begin{equation}
D(\phi) = D_{solvent} + (D_{solid} - D_{solvent})h(\phi)
\label{eq:diff_interp}
\end{equation}

where $D_{solvent}$ is the diffusivity of the drug in the solvent medium, and $D_{solid}$ is assigned a small, non-zero value ($D_{solid} \ll D_{solvent}$) to maintain numerical stability of the diffusion operator while ensuring negligible transport within the solid pill. The interpolation function $h(\phi)$ is chosen as a standard quintic polynomial \cite{yang2024melting}:

\begin{equation}
h(\phi) = \phi^3(10 - 15\phi + 6\phi^2)
\label{eq:smooth_step}
\end{equation}

Next, we define the local dissolution kinetics. Given a set of $S$ excipients with intrinsic dissolution rate constants $\{k^{(1)}, \dots, k^{(S)}\}$, the effective dissolution rate field $k(\bm{x})$ is computed using a linear mixture rule based on the local excipient volume fractions:

\begin{equation}
k(\bm{\gamma}(\bm{x})) = \sum_{s=1}^{S} \gamma_s(\bm{x}) k^{(s)}
\label{eq:diss_rate_mix}
\end{equation}

{\Cref{eq:diss_rate_mix} serves as a continuous interpolation that enables gradient-based optimization of the material assignment. It is not intended to capture microstructural effects such as porosity changes or constituent interactions that may arise in physical multi-material dissolution \cite{moroney2021mathematical}. Its purpose is to map the relaxed excipient fractions to an effective local dissolution rate during intermediate optimization iterates.} Note that while this rate field $k(\bm{x})$ is defined throughout the computational domain, the physical dissolution process is localized to the solid-solvent interface. This localization is explicitly enforced by the $|\nabla \phi|$ term in the governing dissolution forcing function $f_{diss}$ (\Cref{eq:forcing}). Finally, we assume that other material constants including the interface mobililty $(M_{\phi})$, double well potential $(W)$, interface thickness $(\epsilon_t)$, solid mass density $(\rho_s)$, and the saturation concentration $(C_{sat})$ are predefined constants, independent of the excipients and the phase.

With the pill's {geometry}, material distribution, and effective property fields defined, we now proceed to discuss the finite element simulation of the dissolution and diffusion process in the next section.

%----------------------------------%
\subsection{Finite Element Analysis} \label{sec:method_fea}

With the initial phase $\phi(\bm{x}, 0)$ and the spatially varying effective material properties established, the dissolution of the pill is computed by solving the coupled nonlinear governing equations (\Cref{eq:residual_form_phi} and \Cref{eq:residual_form_c}). In particular, we employ standard finite element analysis (FEA) to obtain the spatio-temporal distribution of the phase and concentration fields.

For spatial discretization, the domain $\Omega$ is meshed using four-noded quadrilateral element. The phase field $\phi$ and concentration $C$ are approximated using standard bilinear shape functions. For temporal discretization, we employ an implicit, first-order backward Euler scheme. This yields the following system of semi-discrete residual equations at time step $n+1$:

\begin{equation}
\begin{aligned}
\bm{R}_{\phi} &:= \bm{M} \frac{\bm{\phi}_{n+1} - \bm{\phi}_n}{\Delta t} + \bm{K}_{\phi} \bm{\phi}_{n+1} + \bm{F}_{\phi} = \bm{0} \\
\bm{R}_{c} &:= \bm{M} \frac{\bm{C}_{n+1} - \bm{C}_n}{\Delta t} + \bm{K}_{c} \bm{C}_{n+1} - \bm{F}_{c} = \bm{0}
\end{aligned}
\label{eq:fea_residuals}
\end{equation}

where the mass matrix $(\bm{M})$, stiffness matrices ($\bm{K}_{\phi}$, $\bm{K}_{c}$) and force vectors ($\bm{F}_{\phi}$, $\bm{F}_{c}$) are defined as:

\begin{equation}
\begin{aligned}
\bm{M} &= \int_{\Omega_e} \bm{N}^T \bm{N} \, d\Omega \\
\bm{K}_{\phi} &= M_{\phi} W \epsilon_t^2 \int_{\Omega_e} \bm{B}^T \bm{B} \, d\Omega \\
\bm{F}_{\phi} &= M_{\phi} \int_{\Omega_e} \bm{N}^T \psi'(\phi) \, d\Omega + \frac{1}{\rho_{solid}} \int_{\Omega_e} \bm{N}^T k(C_{sat} - C) |\nabla \phi| \, d\Omega \\
\bm{K}_{c} &= \int_{\Omega_e} \bm{B}^T D(\phi) \bm{B} \, d\Omega \\
\bm{F}_{c} &= \int_{\Omega_e} \bm{N}^T k(C_{sat} - C) |\nabla \phi| \, d\Omega
\end{aligned}
\label{eq:fea_matrices}
\end{equation}

Here, $\bm{N}$ denotes the shape function vector and $\bm{B}$ the gradient matrix. Observe that the stiffness matrix $\bm{K}_{\phi}$, and the force vectors $\bm{F}_{\phi}$ and $\bm{F}_{c}$ render the residuals coupled.

The coupled non-linear system is solved using a Newton-Raphson iterative scheme. The solver is implemented within the JAX \cite{jax2018github} framework and integrated with PETSc \cite{dalcinpazklercosimo2011PetSc} for scalable linear algebra. Crucially, the implementation ensures the entire simulation framework is end-to-end differentiable, facilitating the computation of sensitivities via automatic differentiation \cite{balay2019petsc}.

%--------------------------------------------%

\subsection{Optimization Formulation} \label{sec:method_optimization}

Having established our governing equations and procedures to solve them, the design representation, and the material model we now outline the key components of the design optimization framework.

%----------------------------------%
\subsubsection{Design Variables}
\label{sec:method_optimization_designVar}

Recall that the optimization problem concurrently determines two aspects of the design: (i) the {geometry} of the pill, governed by the geometric parameters of the supershape, and (ii) the distribution of the excipients, governed by the weights of the coordinate-based neural network. The complete design state is therefore defined by the aggregate set of variables $\mathcal{D} = \{ \bm{w}, \bm{\tilde{\zeta}} \}$.

Furthermore, while the neural network weights $\bm{w}$ are unconstrained variables, the parameters of the supershape are bounded to ensure the generated shapes remain physically meaningful and within the computational domain. We define the lower and upper bounds for the geometric parameters as $\underline{\bm{\zeta}}$ and $\overline{\bm{\zeta}}$, respectively:

\begin{equation}
\underline{\bm{\zeta}} = [\underline{c}_x, \underline{c}_y, \underline{\theta}, \underline{a}, \underline{b}, \underline{n}, \underline{m}]^\top, \quad
\overline{\bm{\zeta}} = [\overline{c}_x, \overline{c}_y, \overline{\theta}, \overline{a}, \overline{b}, \overline{n}, \overline{m}]^\top
\label{eq:geo_bounds}
\end{equation}

To incorporate these box constraints into an unconstrained gradient-based optimization framework, we perform the optimization over a set of latent, unconstrained variables $\tilde{\bm{\zeta}} \in \mathbb{R}^7$. The physical design variables are then recovered via a sigmoid projection $\sigma(\cdot)$ and rescaling to the prescribed bounded intervals:

\begin{equation}
\zeta_i = \underline{\zeta}_i + (\overline{\zeta}_i - \underline{\zeta}_i) \sigma(\tilde{\zeta}_i)
\label{eq:sigmoid_transform}
\end{equation}

%----------------------------------%
\subsubsection{Objective}
\label{sec:method_optimization_objective}

The primary goal of the proposed framework is to solve the inverse design problem: determining the optimal pill {geometry} and excipient distribution such that the resulting drug release kinetics match an a priori prescribed target profile. We achieve this by minimizing the discrepancy between the release history of the current design iteration and the target trajectory.

First, we quantify the drug release kinetics by computing the instantaneous rate of mass change of the solid pill. Utilizing the discrete phase-field solution derived from the FEA, the mass evolution rate $\dot{m}$ at time step $t_n$ is approximated as:

\begin{equation}
\dot{m}_n = \frac{\rho_{s}}{\Delta t} \sum_{e=1}^{N_{e}} (\phi_n(\bm{x}_e) - \phi_{n-1}(\bm{x}_e)) v_e
\label{eq:mass_rate}
\end{equation}

where $\rho_{s}$ is the solid density, $\Delta t$ is the time step size, $v_e$ denotes the volume of element $e$, and the summation is performed over all elements in the domain. The optimization problem is then formulated as the minimization of the Mean Squared Error (MSE) between the simulated mass rate profile $\dot{m}$ and the target profile $\dot{m}^*$ over the total simulation time $T$:

\begin{equation}
J = \text{MSE}(\dot{m}, \dot{m}^*) = \frac{1}{N_t} \sum_{n=1}^{N_t} (\dot{m}_n - \dot{m}^*_n)^2
\label{eq:objective_function}
\end{equation}

where $N_t$ is the total number of time steps. This objective function drives the optimizer to update the geometric and material design variables such that the dissolution kinetics align with the desired drug release profile.

%----------------------------------%
\subsubsection{Grayness Constraint}
\label{sec:method_optimization_grayCons}

Recall that while we relaxed $\gamma_s(\bm{x}) \in \{ 0, 1\}$ to $0 \le \gamma_p(\bm{x}) \le 1$ to enable gradient based optimization, physical realizability necessitates a discrete material assignment where $\gamma_s(\bm{x}) \in \{0, 1\}$. To drive the excipient distribution towards this binary state, we impose a global grayness suppression constraint:

\begin{equation}
g_{r} \coloneqq \frac{1}{S n_e} \sum_{e} \sum_{s} \gamma_s (1 - \gamma_s) - \xi \le 0
\label{eq:grayness_constraint}
\end{equation}

where the term $\xi > 0$ is a slack variable governed by a continuation strategy standard in TO. The parameter $\xi$ is initialized with a sufficiently large value, allowing the optimizer to explore the design space with intermediate material mixtures (gray regions). As the optimization proceeds, $\xi$ is reduced toward zero, strictly penalizing intermediate values and forcing the design variables $\gamma_s$ to converge to the binary bounds $\{0, 1\}$.
%----------------------------------%
\subsubsection{Volume Constraint}
\label{sec:method_optimization_volCons}

To satisfy therapeutic and logistical requirements, the design may need to retain a minimum quantity of specific excipients. Towards this, we define the normalized volume deficit $\lambda_s$ for the $s$-th material relative to the total volume of the solid pill as:

\begin{equation}
    \lambda_s \coloneqq \frac{1}{\lambda_s^* V_{solid}} \bigg(\sum_{e=1}^{N_e} \gamma_s(\bm{x}_e) \phi(\bm{x}_e) v_e \bigg) - 1
\end{equation}

where $\lambda_s^*$ is the minimum required volume fraction for excipient $s$, $v_e$ is the element volume, and $V_{solid} = \sum \phi(\bm{x}_e) v_e$ is the total volume of the solid domain. We aggregate the constraints for all $S$ materials using a smooth minimum formulation to define the global constraint $g_{v}$:

\begin{equation}
    g_{v} \coloneqq \frac{1}{\alpha} \log \left( \sum_{s=1}^{S} \exp(-\alpha \lambda_s) \right) \leq 0
    \label{eq:volume_constraint}
\end{equation}

where $\alpha$ (=10 in our experiments) is a sharpness parameter. Satisfying $g_{vol} \leq 0$ ensures that the volume fraction of every constituent satisfies $\lambda_s \geq 0$, thereby meeting the required threshold.

%----------------------------------%

\subsubsection{Optimization Problem}
\label{sec:method_optimization_optProblem}

Collecting the objective, PDEs, grayness and volume constraints the optimization problem can be expressed as:

\begin{subequations}
	\label{eq:optimization_nn_Eqn}
\begin{align}
    \underset{\bm{w}, \bm{\zeta}}{\text{minimize}} \quad & J \\
    \text{subject to} \quad & R_{\phi}(\phi, C) = 0 \\
    & R_{c}(\phi, C) = 0 \\
    & g_{r} \le 0 \\
    & g_v \leq 0
\end{align}
\end{subequations}

We now consider solving the NN-based optimization problem in \Cref{eq:optimization_nn_Eqn}. Neural networks are designed to minimize a loss function using well-known optimization techniques such as Adam procedure (\cite{kingma2014adam,joglekar2024dmf}).
{This is particularly convenient in the present setting because the design variables are the network weights, so the optimization is carried out in a nonlinear implicit design space.}
We therefore convert the constrained minimization problem into a loss function minimization by employing a log-barrier scheme as proposed in \cite{kervadec2022logBarrier, nocedal1999Optimization}. Specifically, the loss function is defined as

\begin{equation}
    \mathcal{L}(\bm{w}) = \frac{J}{J^0} + \psi(g_r) + \psi(g_v)
    \label{eq:lossFunction}
\end{equation}
where, 
\begin{equation}
    \psi_{\tau}(g) = \begin{cases}
    -\frac{1}{\tau} \ln(-g) \; ,\quad g \leq \frac{-1}{\tau^2}\\
    \tau g - \frac{1}{\tau} \ln(\frac{1}{\tau^2}) + \frac{1}{\tau} \; , \quad \text{otherwise}
    \end{cases}
    \label{eq:log_barrier}
\end{equation}

and $J^0$ is the initial objective. The constraint penalty parameter $\tau$ is updated at each iteration $j$ as $\tau = \tau_0 \nu^j$ (where, $\tau_0 = 3$ and $\nu = 1.04$ in our experiments), making the enforcement of the constraint stricter as the optimization progresses. The gradient-based Adam optimizer \cite{kingma2014adam} is used to minimize \Cref{eq:lossFunction}.

%----------------------------------%
\subsection{Sensitivity Analysis}
\label{sec:method_optimization_sensAnalysis}

A critical step in gradient-based {inverse design} is the computation of sensitivities, which are the derivatives of the objective and constraint functions with respect to the design variables \cite{chandrasekhar2023frc,padhy2025toflux, padhy2025tomatoes}. The sensitivity analysis for this work is particularly involved due to the coupling of transient, nonlinear dissolution kinetics with a hybrid design parameterization integrating supershapes and neural networks.

To address this complexity, we construct an end-to-end differentiable pipeline using the automatic differentiation (AD) capabilities of the JAX framework \cite{jax2018github}. This allows us to avoid the laborious and error-prone process of manually deriving complex sensitivity expressions. By implementing the entire forward analysis, from the supershape geometric mapping and neural material distribution to the nonlinear finite element solver, the framework computes the required derivatives to machine precision using reverse-mode AD.

Furthermore, we emphasize on two specific challenges inherent to our physics simulation. The first arises from the use of iterative schemes, such as the Newton-Raphson method, to solve the nonlinear coupled phase-field equations. A naive application of AD would unroll the derivative computation through every solver iteration, a process that is both computationally expensive and memory-intensive. To avoid the prohibitive cost, we apply the Implicit Function Theorem (IFT) \cite{blondel2022ImplicitDifferentiation, padhy2025toflux}, which enables the direct computation of derivatives from the final converged solution, thereby bypassing the need to backpropagate through the iterative solution history.

The second challenge stems from the transient nature of the simulation, which creates a significant memory bottleneck for the adjoint sensitivity method. An adjoint analysis requires access to the state history (concentration and phase fields) from all previous time steps to compute the gradient at the current step, and storing this entire state history is often infeasible for long simulations. We mitigate this issue by employing a checkpointing scheme \cite{wang2009minimal, james2015topology}. This technique stores the system's state at select time steps, and during the reverse-time adjoint solve, states between these checkpoints are recomputed on the fly. This method significantly reduces memory requirements at the cost of a moderate increase in computation time, rendering the analysis of long-duration transient problems feasible.

%----------------------------------%
\subsection{Algorithm}
\label{sec:method_optimization_algorithm}

This section summarizes the optimization framework. The iterative procedure updates the pill {geometry} and the internal excipient distribution to minimize the error between the simulated and target release profiles. The algorithm steps are:

\begin{enumerate}
  \item \textbf{Initialization:} The computational domain $\Omega$ and the finite element mesh are defined. The material constants: specifically the solid density $\rho_s$, potential barrier $W$, interface thickness $\epsilon_t$ and mobility $M_{\phi}$ are assigned, and the target release profile $\dot{m}^*$ is prescribed. The lower and upper bounds for the supershape parameters, $\underline{\boldsymbol{\zeta}}$ and
  $\overline{\boldsymbol{\zeta}}$, are specified (\Cref{sec:method_optimization_designVar}). The geometric design variables $\boldsymbol{\zeta}$ are initialized as the mean of these box constraints. The neural network weight $\mathbf{w}$ are initialized to assign equal volume fractions ($1/S$) to each constituent excipient phase (\Cref{sec:method_designRep_NN}).
  
  \item \textbf{Shape Generation:} In each iteration, the physical geometric
  parameters $\boldsymbol{\zeta}$ are recovered from the latent variables using the sigmoid projection (\Cref{eq:sigmoid_transform}). The supershape radius $R(\varphi)$ is computed (\Cref{eq:supershape_radius}), and the radial distance field $\Phi(\mathbf{x})$ from  \Cref{eq:distance_field_supershape} is
  projected to the smooth phase field $\phi(\mathbf{x})$ (\Cref{eq:density_projection}).

  \item \textbf{Material Distribution:} The spatial coordinates $\mathbf{x}$ are passed through the coordinate-based neural network (\Cref{fig:nn_architecture}) to generate the excipient distribution field $\boldsymbol{\gamma}(\mathbf{x})$, ensuring the partition of unity constraint (\Cref{sec:method_designRep_NN}).

  \item \textbf{Property Mapping:} The phase field $\phi$ and material distribution $\boldsymbol{\gamma}$ are mapped to the finite element mesh to define the spatially varying effective material properties. The local dissolution rate $k(\mathbf{x})$ is computed via the mixture rule (\Cref{eq:diss_rate_mix}), and
  the diffusion coefficient $D(\mathbf{x})$ is interpolated (\Cref{eq:diff_interp}).

  \item \textbf{Finite Element Analysis:} The coupled system of nonlinear transient equations (\Cref{eq:residual_form_c,eq:residual_form_phi}) are solved using standard finite element procedures to simulate the dissolution process (\Cref{sec:method_fea}). The initial phase field $\phi(\mathbf{x}, 0)$ for the simulation is defined by the optimizer output from Step 2. The solution yields the time-dependent phase and concentration fields, which are used to compute the instantaneous mass release rate $\dot{m}$ (\Cref{eq:mass_rate}).

  \item \textbf{Loss and Constraint Computation:} The objective function $J$ is evaluated as the mean squared error between the simulated and target profiles (\Cref{eq:objective_function}). The grayness constraint $g_r$ and volume constraint $g_v$ are computed using \Cref{eq:grayness_constraint,eq:volume_constraint} respectively. The total loss $\mathcal{L}$ is computed by aggregating the objective and the log-barrier penalty term for
  the constraints (\Cref{eq:lossFunction}).

  \item \textbf{Sensitivity Analysis:} The gradients of the total loss with respect to the design variables, $\nabla_{\mathbf{w}} \mathcal{L}$ and $\nabla_{\boldsymbol{\zeta}} \mathcal{L}$, are computed using automatic differentiation (\Cref{sec:method_optimization_sensAnalysis}). This step leverages end-to-end differentiability of the JAX implementation to backpropagate gradients through the transient physics solver.

  \item \textbf{Update:} The design variables are updated using the Adam optimizer. The grayness slack, log-barrier penalty, are updated according to a continuation scheme to gradually enforce the constraints.

  \item \textbf{Termination:} Steps 2 through 8 are repeated until the convergence criteria are met or the maximum number of iterations is reached.
\end{enumerate}

%--------------------------------------------%

%% file: 4_results.tex
\section{Results} 
\label{sec:results}

In this section, we conduct several experiments to illustrate the proposed framework. All experiments are conducted on a MacBook M5 Pro using the JAX library \cite{jax2018github} in Python. Further, we adopt a non-dimensional formulation. Unless otherwise specified, the default parameters for all numerical examples are as follows:

\begin{itemize}
    \item \textbf{Mesh} : A structured bilinear quadrilateral mesh with $75 \times 75$ elements is used. A domain of size $1 \times 1 $ is used.

   \item \textbf{Neural network} : A network with $2$ hidden layers; with $40$ neurons in each layer is used.

   \item \textbf{Constraints} : We enforce a grayness constraint governed by a continuation strategy, where the slack variable $\xi$ is initialized to $2$ and linearly decremented by $5 \times 10^{-2}$ per iteration to a final value of $5 \times 10^{-2}$. Additionally, we impose a minimum volume fraction requirement of $5\%$ ($\lambda_s^* = 5 \times 10^{-2}$) for each excipient.

   \item \textbf{Optimizer} : The ADAM optimizer \cite{kingma2014adam} with a learning rate of $8 \times 10^{-3}$ is used. To improve stability, a gradient clip with a norm threshold of $1$ is used.

    \item \textbf{Convergence} : The optimization is terminated either till a maximum of $100$ iteration or when $ \Delta L \leq 10^{-3}$.

   \item \textbf{Materials} : We adopt a non-dimensional parameter set for all numerical experiments. The specific material properties, including diffusivities, saturation concentration, density, and Allen–Cahn coefficients, are reported in \Cref{table:material_properties}, while the dissolution constants for our excipients are provided in \Cref{table:dissolution_constants}.

\end{itemize}

\begin{table}[h]
\centering
\begin{tabular}{|c|c|}
\hline
\textbf{Parameter} & \textbf{Default value} \\ \hline
$D_{solvent}$ & $10^{-6}$ \\ \hline
$D_{solid}$ & $5 \times 10^{-11}$ \\ \hline
$C_{sat}$ & 1.0 \\ \hline
$\rho_{s}$ & 5.0 \\ \hline
$\epsilon$ & $10^{-4}$ \\ \hline
$W$ & 140 \\ \hline
$M_{\phi}$ & $2 \times 10^{-3}$ \\ \hline
\end{tabular}
\caption{Default material properties.}
\label{table:material_properties}
\end{table}

\begin{table}[h]
\centering
\begin{tabular}{|c|c|c|}
\hline
\textbf{Material} & \textbf{Color} & \textbf{k ($\times 10^{-4}$)} \\ \hline
0 & {\color[HTML]{FF2FA3} Pink} & 0.1 \\ \hline
1 & {\color[HTML]{00C2FF} Blue} & 0.5 \\ \hline
2 & {\color[HTML]{8BE000} Green} & 1.0 \\ \hline
3 & {\color[HTML]{FFC700} Yellow} & 5.0 \\ \hline
4 & White & 0.0 \\ \hline
\end{tabular}
\caption{Multi-material dissolution rate constants.}
\label{table:dissolution_constants}
\end{table}

\subsection{Validation of Hypothesis}
\label{sec:results_validationHypothesis}

The central hypothesis of this work posits that the expanded design freedom offered by multi-material {inverse design} is required for tailoring dissolution kinetics to complex, non-monotonic target profiles. In this section, we validate this hypothesis by comparing the performance of single-material geometric optimization against the proposed multi-material framework.

\begin{figure}[!h]
  \centering
  \includegraphics[scale=0.5]{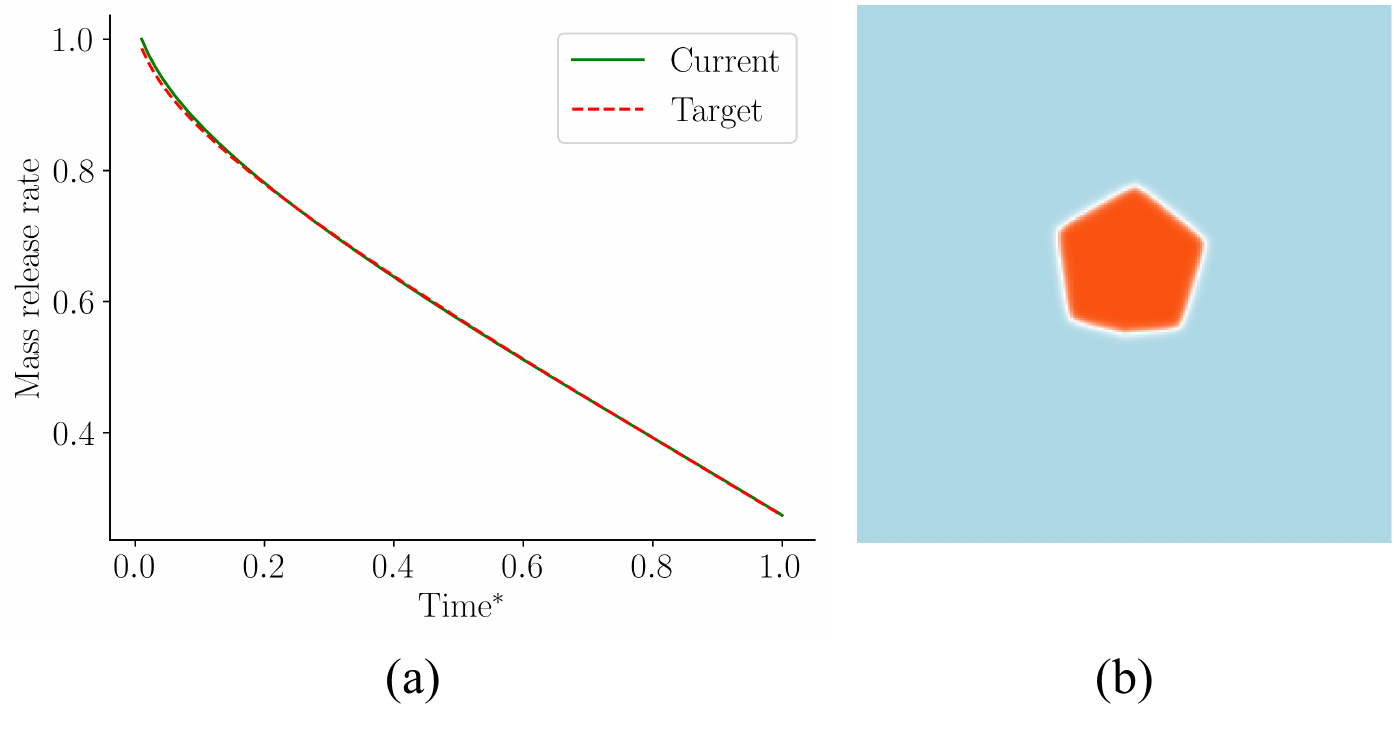}
  \caption{ (a) Comparison of the mass release rate, where the optimized release profile (solid green) matches the monotonic target release (dashed red). (b) The corresponding optimized geometry.}
\label{fig:fig_results_singleMaterial}
\end{figure}

We begin by establishing a baseline for a simple release scenario. We prescribe a simple, approximately monotonic target release curve (see \Cref{fig:fig_results_singleMaterial}(a)). For this case, we restrict the design space to a single excipient material with a dissolution rate constant of $k=2 \times 10^{-4}$. We then optimize the phase distribution as dictated by the supershape parameters to match the target profile. As illustrated in \Cref{fig:fig_results_singleMaterial}, the optimizer successfully converges to a geometry that matched the target release curve, demonstrating that optimization of phase distribution is sufficient for simple release behaviors.

\begin{figure}[!h]
  \centering
  \includegraphics[scale=0.24]{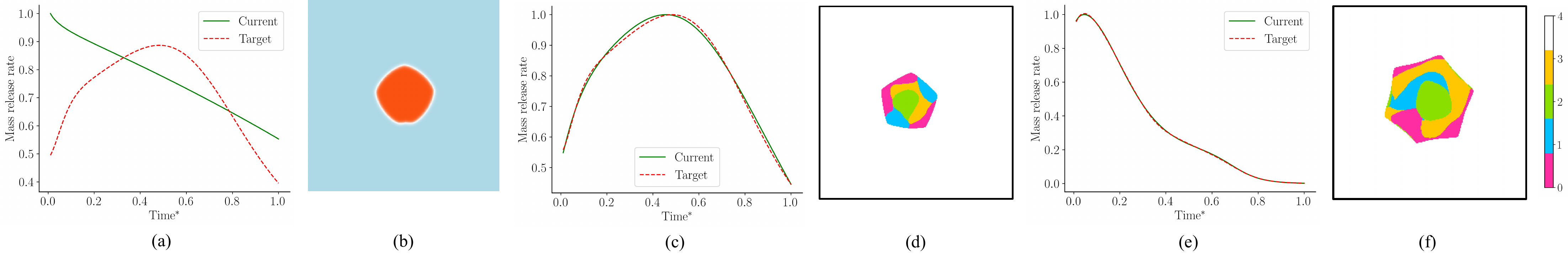}
  \caption{(a) Comparison of mass release rates, demonstrating the failure of the single-material design (solid green) to match the non-monotonic target release curve (dashed red). (b) The corresponding optimized {geometry}. (c) Performance of the multi-material design, where the optimized release profile matches the complex target trajectory. (d) The corresponding multi-material design.{(e) Performance of a target profile with burst-dominated release, gradual decay, and a mild late-time increase. (f) The corresponding optimized multi-material design.}}
\label{fig:fig_results_multiMatNeed}
\end{figure}

However, advanced drug delivery applications often demand complex, modulated release profiles that differ from standard monotonic decay kinetics. To investigate this, we consider a non-monotonic target curve characterized by an initial slow release rate, followed by a rapid acceleration in release, and finally a sharp decay (see \Cref{fig:fig_results_multiMatNeed}(a)). We attempt to match this target using the single-material strategy, optimizing only the phase distribution via the supershape parameters. As shown in \Cref{fig:fig_results_multiMatNeed}((a) and (b)), the optimization fails to capture the transition from slow onset to rapid release. This result highlights the inherent limitations of a single-material design space, where the dissolution rate mainly depends on the evolving surface area of the pill.

To circumvent this limitation, we introduce multiple excipient materials. We concurrently optimize the phase distribution (via supershape parameters) and the internal material distribution (via the neural network representation). Spatially varying the excipient composition with distinct dissolution constants (see \Cref{table:dissolution_constants}) expands the achievable release kinetics beyond those dictated by exposed surface area alone. Consequently, the multi-material design matches the complex target profile (\Cref{fig:fig_results_multiMatNeed} (c) and (d)). {Quantitatively, the single-material design (\Cref{fig:fig_results_multiMatNeed} (a)) yields a normalized MSE of $3.01 \times 10^{-2}$ (normalized RMSE of $17.4\%$), indicating its limited ability to match the non-monotonic target. In contrast, the multi-material design (\Cref{fig:fig_results_multiMatNeed} (c)) achieves a normalized MSE of $9.52 \times 10^{-5}$ (normalized RMSE of $0.98\%$).} This result validates our hypothesis that the expanded design freedom of multi-material TO is essential for realizing complex release behaviors. {Furthermore, the optimized material distribution in \Cref{fig:fig_results_multiMatNeed}(c, d) provides physical insight into how the optimizer exploits multi-material design freedom. The non-monotonic target demands three phases: slow initial release, rapid acceleration, and sharp decay. To achieve this, the optimizer places the slowest-dissolving excipient (Pink) at the pill periphery, suppressing early release while the solvent interacts primarily with this resistant outer shell. The fastest-dissolving excipient (Yellow) is concentrated in the interior; once the outer shell erodes, a large surface area of this rapidly dissolving material is exposed, producing the sharp peak in the release curve near $T^* \approx 0.4$. The subsequent decay follows as this region is consumed and the remaining solid volume diminishes. In contrast, the single-material design (\Cref{fig:fig_results_multiMatNeed}(a,b)) can only modulate release through the evolving surface area, which for convex geometries decreases monotonically. As a result, it cannot reproduce the non-monotonic profile.}

{To further illustrate the scope of the method, we include an additional representative target profile beyond the non-monotonic example discussed earlier. This example exhibits a burst-dominated release profile, with a high initial release rate followed by gradual decay and a mild late-time increase. The corresponding optimized design and release curve are illustrated in \Cref{fig:fig_results_multiMatNeed}(e,f); the optimizer yields a normalized MSE of $6.44 \times 10^{-6}$ (normalized RMSE of 0.25\%), indicating excellent agreement with the prescribed target. Similar orders of normalized MSE values were observed across the other multi-material examples.\\
While these results demonstrate that the framework can match a range of qualitatively distinct release behaviors, the achievable set is not unbounded. As the present discussion is restricted to the dissolution-dominated regime studied here under the idealized sink-condition setting. In these examples, the computed response corresponds to net mass transfer from the solid to the solvent, so the release rate is nonnegative and, under full dissolution, decays to zero as the remaining solid is consumed. Hence, target profiles that require sustained growth near the end of the release window, nonzero terminal release, or negative release are outside the achievable set of the current model. Very sharp discontinuities or narrow pulses are also difficult to reproduce exactly because the diffuse-interface formulation and implicit time stepping smooth the release dynamics.}

{More broadly, the reachable set of release profiles is governed jointly by geometry and material distribution. Geometry controls the exposed interface area over time, while the constituent layout controls which local dissolution rates are encountered by that interface. A precise analytical characterization of this reachable set is beyond the scope of the present work, because the interface area evolves nonlinearly during dissolution and depends jointly on geometry and material arrangement. The grayness and volume constraints further restrict this set by enforcing discrete material assignments and limiting the admissible proportions of fast- and slow-dissolving excipients.}

\subsection{Effect of Initialization}
\label{sec:results_effectInitialization}

Having demonstrated the necessity of multi-material {inverse design}, we now investigate the influence of design initialization on optimization result. Given the inherent non-convexity and non-linearity of the coupled dissolution-transport problem, we anticipate that the optimization landscape contains multiple distinct local minima.

\begin{figure}[!h]
  \centering
  \includegraphics[scale=0.45]{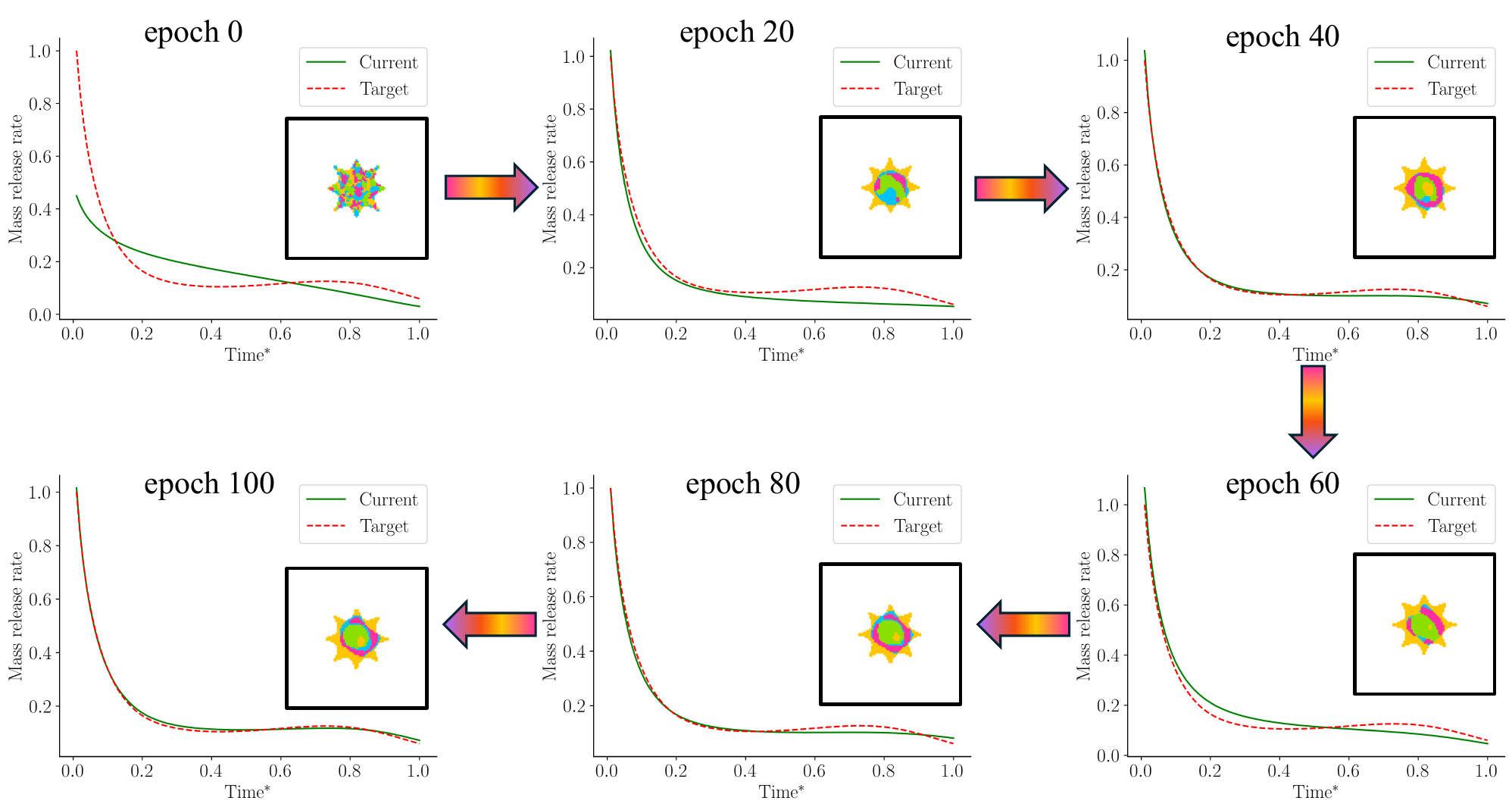}
  \caption{Convergence history of the multi-material {inverse design}. The sequence (top to bottom) shows the update of the design and release curve from initialization (Epoch 0) to the final optimized state (Epoch 100). }
\label{fig:fig_results_convergence}
\end{figure}

To explore this, we first initialize the phase distribution using a supershape parametrization corresponding to a highly lobed geometry (referred to as the ``spike'' design). We prescribe the complex target release profile shown in \Cref{fig:fig_results_convergence}. The specific parameter bounds for this initialization are detailed in \Cref{table:supershape_bounds}.

\begin{figure}[!h]
  \centering
  \includegraphics[scale=0.45]{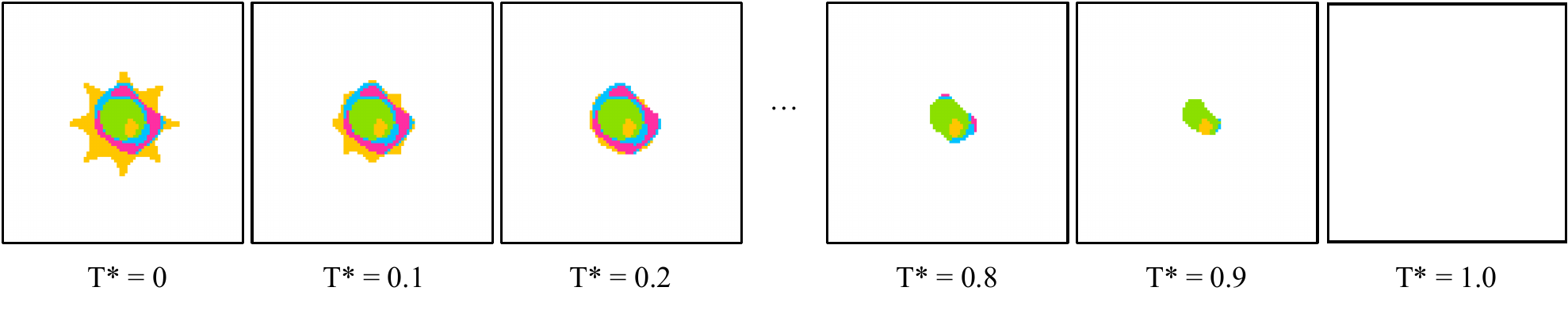}
  \caption{Dissolution of the optimized pill.}
\label{fig:fig_results_ConvergencePillDissolution}
\end{figure}

Consistent with previous results, the multi-material {inverse design} successfully converges to a design that matches the target release curve. \Cref{fig:fig_results_convergence} illustrates the convergence behavior. The initial design exhibits poor performance with a significant deviation from the target release curve. However, subsequent iterations substantially modify both the supershape parameters (phase distribution) and the neural network weights (material distribution), achieving a good fit by the $100th$ iteration. Crucially, the final design satisfies the grayness constraint, ensuring discrete material allocation without nonphysical mixing, and satisfies the minimum material fraction constraint. The optimization converges in approximately $30$ minutes, averaging $10$ Newton-Raphson iterations per time step (over 100 total time steps) for the combined forward and backward computation. Finally, the temporal evolution of the optimized ``spike'' design is depicted in \Cref{fig:fig_results_ConvergencePillDissolution}, visualizing the progressive dissolution of the multi-material interface.

While the ``spike'' initialization yields a dissolution rate that closely matches the target, the resulting {design} (\Cref{fig:fig_results_convergence}) may pose challenges regarding fabrication and patient ingestibility due to its sharp geometric features. To obtain designs more suitable for practical application, we explore alternative initializations—specifically ``circle'' and ``sunflower'' geometries by imposing constraints on the supershape parameters. By bounding the extent of the shape parameters (see \Cref{table:supershape_bounds}), we constrain the optimizer to explore regions of the design space characterized by smoother designs favorable for ingestion.

\begin{table}[h]
    \centering
    \begin{tabular}{l c c c}
        \hline
        \textbf{Design Type} & \textbf{Curvature ($n$)} & \textbf{Lobes ($m$)} & \textbf{Scale ($s$)} \\ \hline
        Spike & $[0.5, 2.0]$ & $[5, 11]$ & $[0.1, 0.4]$ \\
        Circle & $[1.67, 2.0]$ & $[1, 3]$ & $[0.1, 0.4]$ \\
        Sunflower & $[2.5, 4.0]$ & $[10, 14]$ & $[0.1, 0.4]$ \\ \hline
    \end{tabular}
    \caption{Supershape parameter bounds for different design initializations. Here, $n$ governs the curvature and $m$ denotes the rotational symmetry/number of lobes.}
    \label{table:supershape_bounds}
\end{table}

\begin{figure}[!h]
  \centering
  \includegraphics[scale=0.3]{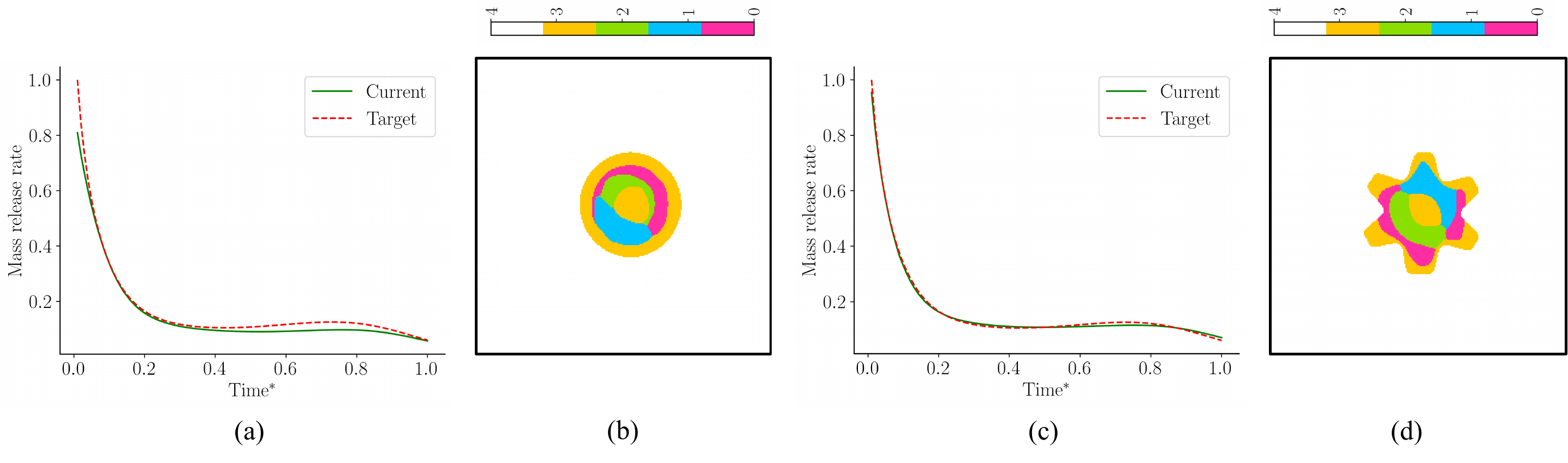}
  \caption{Optimization results for ``circle'' and ``sunflower'' initializations. (a, c) Release profiles demonstrate that the optimized designs accurately track the target profile. (b, d) The corresponding topologies.}
\label{fig:fig_results_initalizations}
\end{figure}

Despite these geometric restrictions, the optimizer successfully matches the target release curve for both the circle \cref{fig:fig_results_initalizations}((a) and (b)) and sunflower initializations \cref{fig:fig_results_initalizations}((c) and (d)). This non-uniqueness demonstrates that the proposed framework is robust, capable of identifying various distinct local optima that yield equivalent functional performance (release profiles). This flexibility allows designers to select optimal configurations based on additional criteria, such as manufacturability or ingestibility, without compromising the release kinetics.

\subsection{Case Study: Effect of Pill Degradation}

To demonstrate the framework's capability in incorporating environmental parameters, we examine a design scenario based on the degradation of pharmaceuticals. For instance, we draw motivation from a critical logistical challenge in long-duration spaceflight: the in-situ fabrication of therapeutics using degraded feedstock. Prolonged exposure to the space environment induces chemical instability in stored pharmaceuticals. For instance, data \cite{du2011evaluation} from the ISS indicates that Clavulanate formulations exhibit a loss of dissolution performance; specifically, samples stored for 880 days show a dissolution efficacy ($Q$-value) of only $4.4\%$. 

\begin{figure}[!h]
  \centering
  \includegraphics[scale=0.3]{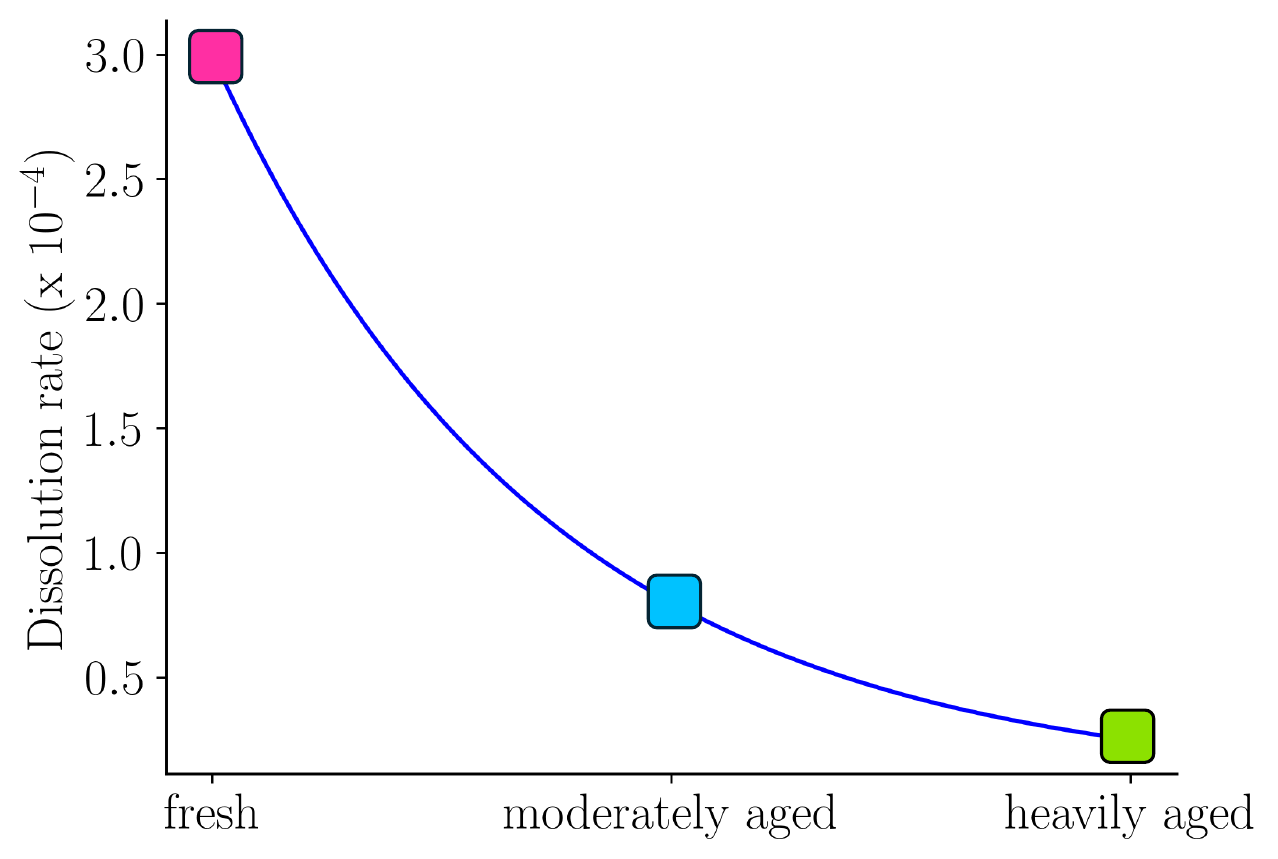}
  \caption{Dissolution rate constants plotted for three available feedstock grades—fresh, moderately aged, and heavily aged.}
\label{fig:fig_results_issExcipients}
\end{figure}

Towards this, we formulate a simplified {inverse design} problem that addresses two concurrent challenges: (1) the degradation of stored feed-stock's dissolution kinetics, and (2) the logistical constraints of limited resupply. We assume an inventory consisting of heavily aged, moderately aged and fresh excipients. The non-dimensional excipients for these three supplies are shown in \Cref{fig:fig_results_issExcipients}.

\begin{figure}[!h]
  \centering
  \includegraphics[scale=0.6]{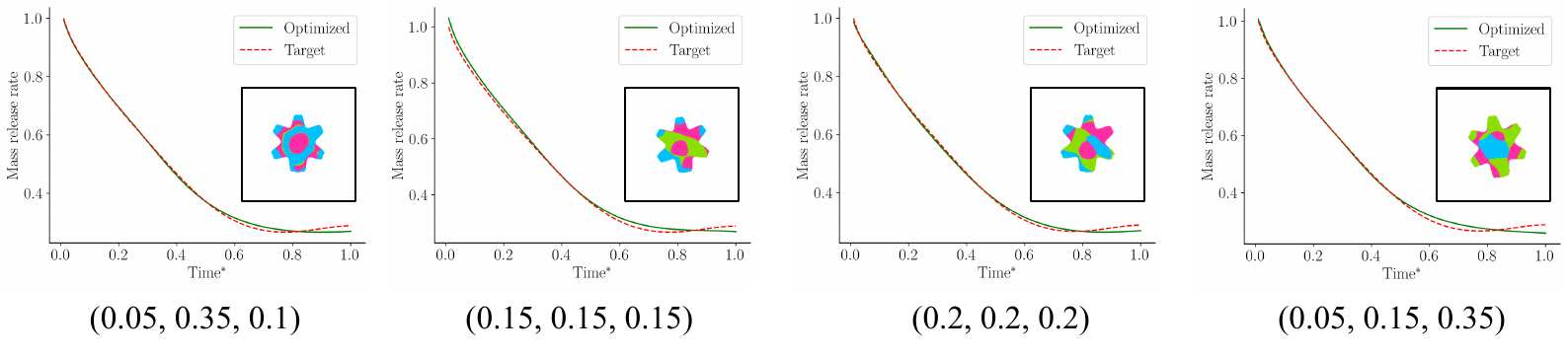}
  \caption{Optimized for varying minimum volume fraction constraints $(\lambda^*_{\text{fresh}}, \lambda^*_{\text{moderate}}, \lambda^*_{\text{aged}})$.}
\label{fig:fig_results_issDesigns}
\end{figure}

We now apply the proposed framework under different minimum volume utilization requirements defined by $  (\lambda^*_{\text{fresh}}, \lambda^*_{\text{moderate}}, \lambda^*_{\text{aged}})$. For example, we consider four cases for different minimum volume usage requirement. The obtained release curve along with the material distribution is shown in \Cref{fig:fig_results_issDesigns}. We observe that we are able to closely match the target dissolution profile by varying the excipient distribution. Notably, as expected, as the volume fraction constraint of heavily aged material increases, the optimizer adapts by allocating a greater volume fraction to the degraded excipient (green). This experiment showcases the utility of our framework to design pills on demand taking into account environmental factors. This case study serves to validate the framework's utility in ensuring therapeutic robustness under conditions of material uncertainty and logistical scarcity assuring medication reliability in remote, resource-constrained environments. {Furthermore, this case study is intended as an illustrative proof-of-concept for incorporating degraded-material constraints into the inverse-design framework, rather than as a mechanistic model of the underlying physicochemical degradation pathways.}

%% file: 5_conclusions.tex
\section{Conclusions} 
\label{sec:conclusions}

This work introduces a computational framework for the co-optimization of the {geometry} and excipient material distribution of polypills to achieve a specified drug release profile. In particular, we employ a hybrid design representation that couples supershapes with a coordinate-based neural network. This formulation parameterizes the pill's {geometry} to ensure a connected domain while simultaneously representing the spatially heterogeneous distribution of excipients with varying dissolution kinetics. We integrate these differentiable representations into a multiphysics environment governed by a modified Allen-Cahn phase field dissolution equation and Fickian diffusion for the simultaneous optimization of pill shape and material composition. Numerical experiments validate our central hypothesis: the co-design yields dosage forms with release profiles that match complex target kinetics.

The present study offers several avenues for future research. First, the dissolution analysis included only a single active ingredient and assumed constant material parameters such as diffusivity and mobility. A more complex model would involve considering multiple active ingredients, the effects of the active ingredient on dissolution kinetics, and environmental factors such as temperature, and the solvent's chemical reactivity. {Furthermore, extending the framework toward \textit{in vivo} relevance would require modeling the gastrointestinal environment, including finite fluid volumes, local concentration buildup, absorptive transport across the intestinal wall, spatially varying pH, mucus-associated transport barriers, and fluid-motion effects such as peristaltic advection \cite{abuhelwa2017food,macadam1993effect,sitovs2025dissolution}. } Second, the material properties used in this work were primarily for illustration. The practical implementation of this framework will require the use of comprehensive, experimentally validated material datasets that account for pH-dependency, complex excipient-drug interactions, and factors such as chemical stability, storage shelf-life, and material compatibility. Third, while the optimization formulation utilized grayness constraints to ensure physical validity, it did not explicitly account for additive manufacturing constraints such as minimum feature size or print resolution. Incorporating manufacturing filters and extending the framework to 3D supershapes or fully volumetric density-based parameterizations will be essential to expand design freedom while ensuring connectivity and physical realizability. {However, a practical 3D extension will likely require dedicated acceleration strategies, such as reduced-order or surrogate models \cite{xiao2024primal} and GPU-based implementations \cite{macklin2022warp}, to keep the transient nonlinear multiphysics optimization computationally tractable.} Fourth, the transient and nonlinear governing equations often exhibited numerical stiffness rendering the  simulations computationally demanding. Consequently, the development of specialized algorithms for nonlinear transient inverse design remains a critical challenge \cite{alexandersen2025large, appel2025one}. Fifth, while we considered a deterministic design scenario, incorporating manufacturing, environmental and material uncertainty remains a significant interest. Sixth, while we considered a single scale designs, advances in additive manufacturing offer the ability to fabricate multi-scale structures that permit greater customization. Seventh, future work should integrate patient-specific factors to model the release curve, fully realizing the potential of personalized medication. Finally, experimental validation of our optimized designs using 3D-printed prototypes is a crucial area for future investigation to verify the computational predictions.